\documentclass[preprint,aps]{revtex4}
\newcommand{\be}{\begin{equation}}
\newcommand{\ee}{\end{equation}}
\newcommand{\bea}{\begin{eqnarray}}
\newcommand{\eea}{\end{eqnarray}}
\newcommand{\de}{\partial}

\usepackage{graphicx}
\usepackage{epsfig}
\usepackage{amssymb}
\usepackage{amsfonts}
\usepackage{comment}
\usepackage{bm}
\usepackage{graphicx}
\usepackage{verbatim}
\usepackage{dsfont}
\begin{document}
\title{Exact out-of-equilibrium central spin dynamics from integrability}

\author{Davide Fioretto$^{1,2}$,  Jean-S\'ebastien Caux$^2$, Vladimir Gritsev$^{1,2}$}
\address{$^1$ Department of Physics, University of
Fribourg, Chemin du Mus\'ee 3, 1700 Fribourg, Switzerland\\
$^2$ Institute for Theoretical Physics, Universiteit van Amsterdam, Science Park 904,
Postbus 94485, 1098 XH Amsterdam, The Netherlands}
\begin{abstract}
We consider a Gaudin magnet (central spin model) with a time-dependent exchange couplings. We explicitly show that the Schr\"odinger equation is analytically solvable in terms of generalized hypergeometric functions for particular choices of the time dependence of the coupling constants. Our method establishes a new link between this system and the $SU(2)$ Wess-Zumino-Witten model, and sheds new light on the implications of integrability in out-of-equilibrium quantum physics. As an application, a driven four-spin system is studied in detail.
\end{abstract}
\pacs{}

\maketitle

\section{Introduction}
The problem of describing the coherent out-of-equilibrium evolution of driven many-body quantum systems has attracted a great deal of attention in recent years. This interest was spurred by the recent advances in cold atoms and semiconductor physics, which made experimental observations possible. The attention of the community has mostly been devoted to the investigation of two limiting cases: the quench regime \cite{quench_Calabrese2006}, where the variation of the parameters of the Hamiltonian is very fast with respect to all the other time scales of the problem, and the adiabatic regime, where it is slow (see \cite{RMP} for a recent overview). Outside of these two extreme situations, very little is known, either analytically or numerically. This is unfortunate, because this driven regime has the most potential for novel physics.\\

There are natural obstacles for direct studies of non-equilibrium quantum many-body systems: only a  few solutions of the single-particle Schr\"{o}dinger equation with time-dependent parameters are known even in the single-particle case. The situation is even worse in the many-body case. While integrable many-body systems provide considerable insight into equilibrium physics in one dimension, their non-equilibrium behavior is still difficult to analyze because of the complexity of their solution. Numerical treatments of time-dependent systems (like e.g. by the time-dependent density-matrix renormalization group) are limited by the quantum entanglement which grows while the system evolves in time starting from some initial state \cite{ent_Calabrese,ent_DeChiara,ent_Lauchli,ent_Alba}. To our knowledge there is a only a single subclass of systems where the full time dependence of parameters can be kept to some extent and where the dynamics can be understood in its full complexity. The dynamics of these systems can be mapped to the  dynamics of a different systems which have no explicit time-dependence of parameters by an appropriate transformation of the coordinates, time, and wave-functions 
\cite{Lewis,Perelomov,Kagan,Pitaevskii,Minguzzi,Gritsev}
. While this class of models is limited, it provides a clue about certain interesting and fundamental dynamical effects, like e.g. dynamical fermionization, and moreover it does not rely on integrability of the time-independent model. 

To go beyond this class of models some new ideas are needed. Here we make an effort in this direction by suggesting to use a fact that the wave functions of a broad class of many-body quantum-mechanical models can be represented in terms of the correlation functions of some {\it field theories} with known properties. These connections were discovered and used in the context of the quantum Hall effect, where the quantum wave functions can be related to conformal blocks of the two-dimensional conformal field theories (CFT) \cite{Moore}. Interestingly, the wave functions of some integrable spin models can also be related to the correlators of certain CFT's \cite{Sierra,Galitski}
. Here we extend and use these observations further to study non-equilibrium dynamics of those spin models. Since these spin models belong to a broader class of a so-called Gaudin systems, our observations can be applied to that class as well. The central ingredient of our approach here is the fact that the conformal blocks of the 2D Wess-Zumino-Witten model are solutions of the Knizhnik-Zamolodchikov equation. For a broader class of CFT's (without internal symmetry) these equations should be replaced by the Belavin-Polyakov-Zamolodchikov system.  We believe that the approach we explore here can be generalized further for systems with a more general matrix-product like structure of the wave functions.

Implementing the above ideas concretely, we investigate a model of $N$ spin$-1/2$ degrees of freedom coupled by time-dependent exchange parameters $J_i (t)$, 
\be\label{central-spin-Ham}
H_{CS}(t)=\sum_{i=1}^{N-1}J_{i}(t){\bf S}_{0}\cdot {\bf S}_{i}.
\ee
Label $0$ refers to the `central spin' which is coupled to the $N-1$ other (mutually uncoupled) spins. For time-independent couplings, this model is known as the central spin model, a Gaudin magnet \cite{GaudinBOOK,Amico,Ortiz}. Crucially, this Hamiltonian is directly relevant to experiments in quantum dots \cite{ReviewQD,Imamoglu} and Nitrogen vacancy (NV) centers in diamond \cite{Lukin}, in which time-dependent couplings are intrinsic to the experimental protocols (respectively via time-dependent gate voltages and external electromagnetic fields). The model (\ref{central-spin-Ham}) is only one of the broad class of models where dynamics can be treated using our method here. Other Gaudin-type models can be directly studied in a similar way. 
 
The aim of this article is therefore twofold. On the one hand, we identify a time-dependent protocol for which it is possible to obtain analytical information (i.e. the exact many-body wavefunction) for a class of Hamiltonians related to the (\ref{central-spin-Ham}). While the requirements of this protocol are restrictive, they nonetheless allow to go beyond the adiabatic or sudden approximation. On the other hand, our technique points to an intriguing link between the time dependent central spin Hamiltonian and the Wess-Zumino-Witten (WZW) model, a well-known conformal field theory (CFT), opening the door to further applications of CFT techniques to driven nonequilibrium physics.

We note that here we will restrict our interests to the dynamics of the total spin-singlet subspace, ${\bf S^2}=0$. This is a good starting point because (see \cite{DFS1,DFS2} and the recent review \cite{Lidar}) this subspace plays a crucial role in quantum information theory: the decoherence-free dynamics naturally occurs in this subspace, while the qubits can be encoded into its basis states.

\section{Main Results}
We begin from the fact that the conformal blocks of the Wess-Zumino-Witten model~\cite{DiFrancesco} satisfy the Knizhnik-Zamolodchikov equations~\cite{KZ}. For a $SU(2)_k$ WZW model, these equations read
\be
\left[k \frac {\de} {\de z_i}-\sum_{j\ne i} \frac{\mathbf{S}_i \cdot \mathbf{S}_j}{z_i-z_j}\right] \Psi_N(z_0,\ldots,z_{N-1})=0,  \label{KZ}
\ee
where $\Psi_N(z_0,\ldots,z_{N-1})= \langle \varphi(z_0)\ldots \varphi(z_{N-1}) \rangle$ is the N-point holomorphic conformal block of primary field $\varphi$, while $k$ is a number known as the level of the Kac-Moody algebra. If $k$ is a positive integer, the WZW model is a rational CFT. Interestingly, there exist integral representations of solutions to the KZ equations that can be analytically continued to any nonzero complex $k$ \cite{Babujian,Schechtman}.\\ 

Choosing $k=i v$ where $v\in \mathds{R}$ and considering the ansatz $\psi_{N}(t)=\Psi_{N}(z_{0}(t),\ldots,z_{N-1}(t))$ for a many-body wavefunction, we see that $\psi_N(t)$ can in fact be reinterpreted as a time-dependent Schr{\"o}dinger equation $i \hbar \dot{\psi}_{N}(t)=H(t)\psi_{N}(t)$ with Hamiltonian
\be
H(t)=\sum_{i=0}^{N-1} \sum_{j \ne i} \frac{\hbar\,\dot{z}_i(t)}{v} \frac{\mathbf{S}_i \cdot \mathbf{S}_j}{z_i(t)-z_j(t)} .
\ee
Therefore, if $z_0(t)$ is chosen to be the sole time-dependent parameter, $\psi_N(t)=\Psi_N(z_0(t),z_1,\ldots,z_{N-1})$ solves the time-dependent Schr{\"o}dinger equation for Hamiltonian (\ref{central-spin-Ham}) with couplings
\be
J_{i}(t)= \frac{\hbar \,\dot{z}_0(t)}{v\,(z_0(t)-z_i)} . \label{coupling}
\ee
It is shown in Appendix \ref{app_wzw}  that this choice of time-dependent parameters $z_j(t)$ ($j=0,\ldots, N-1$) is uniquely dictated by the form of (\ref{central-spin-Ham}). 
Notice that the hermiticity of the Hamiltonian forces all the $z_i$ to be on the same line in the complex plane (for example, we can take them to be all real). \\

Let us emphasize the main features of our approach.
First of all, the main ingredient for an explicit solution of the time-dependent Schr{\"o}dinger equation is a solution of the Knizhnik-Zamolodchikov equations that can be analytically continued to imaginary k. For small systems, this can be done explicitly, using standard CFT techniques. For larger systems, we can rely on a class of integral representations. Quite interestingly,  these representations rely crucially on the integrability of the time-independent Hamiltonian, i. e. the  off-shell Bethe equations. Therefore, the solubility of the time-dependent Schr{\"o}dinger equation seems to be a signature of the underlying integrability of the model that survives also when the couplings are time dependent. Indeed, this interpretation is confirmed by the fact that, as we will discuss later on, the solvability of the time-dependent Schr{\"o}dinger equation is not  a special feature of the central spin model: our arguments apply also to the broader class of XXZ Gaudin magnets. Moreover, our results establish a new connection between the $SU(2)$ WZW model (admittedly, for the quite unusual imaginary $k$ case) and the time-dependent central spin Hamiltonian. It is worth to note here that the $SU(2)$ WZW model is known to be related to integrable\cite{Sierra} and non integrable\cite{Nielsen1,Nielsen2} time-independent spin Hamiltonians. Finally, it is important to stress that-by construction- this approach works only if the time dependence of the $J_i(t)$ is finely tuned: essentially, the time evolution is ``geometric", i.e. $\int dt H(t)$ can be written as a curvilinear integral in the space of the $z_j$'s.\\

 Our paper is organized as follows. First of all, in Sec. \ref{sec_four_spins} we provide a detailed analysis of a simple system of four spins. Thanks to the connection between the central spin Hamiltonian  and  the WZW model, we are able to analyze the time of evolution of the subspace of zero total spin in terms of hypergeometric functions. In this way, we can see our approach explicitly in action and understand some mathematical property of our solution (i.e. completeness and non triviality). Therefore, in Sec. \ref{sec_general}, we move to a more general setting: a N particle XXZ Gaudin magnet with time dependent couplings (\ref{coupling} ). Here, we take advantage of an integral representation of the solution of the (generalized)  Knizhnik-Zamolodchikov equations to provide an integral representation for the time dependent many body wavefunction. While this representation is not (yet) amenable to an quantitative evaluation, it allows us to consider two interesting situations: the adiabatic and the semiclassical limit, thus gaining insight on the completeness of our solutions (sec. \ref{sub_adiabatic}) .
  Finally, we present our conclusions in Sec. \ref{sec_conclusions}, while some of the more technical details are discussed in the  appendices.

\section{A simple example: a four spins system.} 
 \label{sec_four_spins}
 The class of Hamiltonians under consideration has a quite specific time-dependent coupling constant (\ref{coupling}). Moreover, as we will see, in the general case, while it is possible to write down an integral representation for the wavefunction, it is not easy to extract physical predictions from it. The reader could wonder if this class of Hamiltonians can be solved only because their physics is trivial or if, instead, we can expect some interesting phenomenology that might motivate a further investigation of these systems. In this section, we want to address this point by studying one quite simple representative of this class of Hamiltonians: a central spin Hamiltonian with four constituents,
\be
H (t) =\frac 1 v \sum_{i=1}^3 \frac{\dot{z}_0(t)}{z_0(t)-z_i} \mathbf{S}_0\cdot \mathbf{S}_i . \label{example_cs}
\ee
Since the total spin is conserved by the time evolution, we can restrict ourselves to the subspace with constant $\mathbf{S}^2=\left(\sum_i {\bf S}_i\right)^2$. In the following, we would like to show that, indeed, the WZW correlators provide solutions that describe the whole zero spin subspace.\\
The computation of the four point conformal blocks $\Psi_4(z_0,z_1,z_2,z_3 )$  of the WZW model is a standard exercise of CFT (see \cite{DiFrancesco}). The detailed calculation is reported in Appendix \ref{app_four}, where $\Psi_4(z_0,z_1,z_2,z_3 )$ is expressed in terms of the standard hypergeometric functions $\,_2 F_1(a,b,c,x)$\cite{Whittaker}. We introduce the parametrization $  \Psi_4(z_0,z_1,z_2,z_3)=[(z_0-z_3)(z_1-z_2)]^{-\frac{3} {4 k}} f(x) $, where $f(x)$ is a function of the anharmonic ratio $x=\frac{(z_0-z_1)(z_2-z_3)}{(z_0-z_3)(z_2-z_1)}$ and  expand $f(x)=[x(1-x)]^{-\frac 3 {4 k}}\sum_{i=1}^2 G_i(x) |v_i\rangle$  on a basis of the $S^2=0$ subspace given by the two states
\bea
&& \!\!\!\!\!  \!\!\!\!\!  \!\!\!\!\!  \!\!\!\!\!  \!\!\!\!\!  \!\!\!\!\! |v_1\rangle=\left(\frac{|+-\rangle-|-+\rangle}{\sqrt{2}}\right) \otimes \left(\frac{ |+-\rangle-|-+\rangle }{\sqrt{2}}\right) ,\label{basis}\\
&&  \!\!\!\!\!  \!\!\!\!\! \!\!\!\!\!  \!\!\!\!\!  \!\!\!\!\!  \!\!\!\!\! |v_2\rangle=\frac 1 {\sqrt{3}}\left[ |++--\rangle+ |--++\rangle +\left(\frac{ |+-\rangle+|-+\rangle }{\sqrt{2}}\right)\otimes \left(\frac{ |+-\rangle+|-+\rangle }{\sqrt{2}}\right) \right] \nonumber .
\eea
We thus can write $G_1(x)=\sum_{i=1}^2 c_i w_i(x)$, where $c_{1,2}$ are constants determined by the initial conditions, while  
\bea
&&w_1(x) =\,_2F_1 \left(-\frac 3 {2k},-\frac 1 {2 k},-\frac 1 k,x\right) ,\\ \nonumber 
&&w_i(x)=(-x)^{\frac{1+k}{k}} \,_2F_1\left(1-\frac{ 1} {2 k},1+\frac{2k} 2+\frac 1 k,x\right) ,
\eea
and $G_2(x)=\frac {1-x} {\sqrt{3} \,x} \left[3 G_1(x) +4\,k\, x\, G'_1(x) \right]$. Therefore, as discussed above, the wavefunction $\psi_4(t)=\Psi_4(z_0(t),z_1,z_2,z_3)$ (with $k=i v$) is a solution of the time-dependent Schr{\"o}dinger equation for Hamiltonian (\ref{example_cs}).  

As an example, let us consider the following protocol. At time t=0, the spins $\mathbf{S}_j$, $j=1,2,3$, are at a distance j from the central spin $\mathbf{S}_0$. Their couplings $J_j$ are taken to be proportional to $ j^{-3}$ (dipolar interaction) or to $\exp\left(-j^2\right)$ (shell model). Subsequently for $t>0$, the coupling constants decrease inverse linearly in time (plus a site-dependent term). We can thus model this situation with $z_0=\omega t$ and $z_j= -j^3$ (dipolar interaction) or $z_j=- \exp\left(j^2\right)$ (shell model), $j=1,2,3$. 

The first thing to analyze is the completeness of the solution, {\it i.e.} if the space spanned by the conformal blocks solution is bidimensional. It is shown in Appendix \ref{app_four} that the absolute value of the determinant of the matrix $M_{i j}(x(t))$ such that 
$G_i(x(t))=\sum_j M_{i j} (x(t)) c_j$ is actually constant and nonvanishing  for $t\in [0,+\infty)$, thus proving that this family of solutions spans the whole subspace of zero total spin. This fact is unrelated to our choice of $z_0(t)$ and $z_i$ and remains true for any parametrization (see Appendix \ref{app_four}).

 As an application, an interesting quantity to look at is the modulus square of the overlaps of the wavefunction with the basis vectors $|v_i  \rangle$ , i.e. $a_i(t)=|\langle v_i |\psi(t)\rangle |^2$ , which are simply computed. We can expect that, if these overlaps are almost constant in time, then the time evolution is essentially trivial. As a signature of the non triviality of the time evolution, we look the crossing of $a_i(t)$, i.e. when $a_i(t)<a_{j}(t)$ for $t<t^\prime$, while $a_i(t)>a_{j}(t)$  $t>t^\prime$: this means that for $t<t^\prime$ the state $i$ is more important than the state $j$, while the opposite is true for $t>t^\prime$.

\begin{figure*}[t]
\begin{minipage}[b]{0.45\linewidth}
\centering
\includegraphics[totalheight=0.2275\textheight]{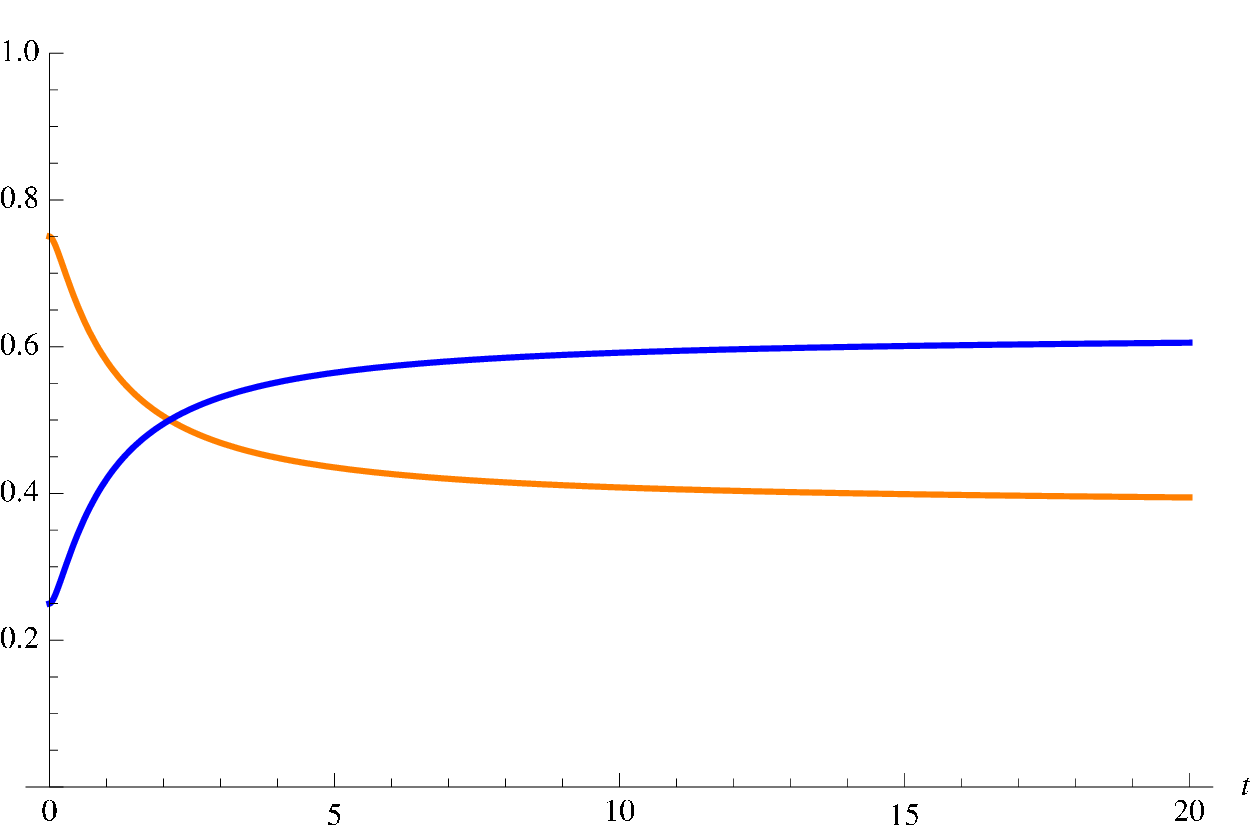}
\end{minipage}
\hspace{0.5cm}
\begin{minipage}[b]{0.45\linewidth}
\centering
\includegraphics[totalheight=0.2275\textheight]{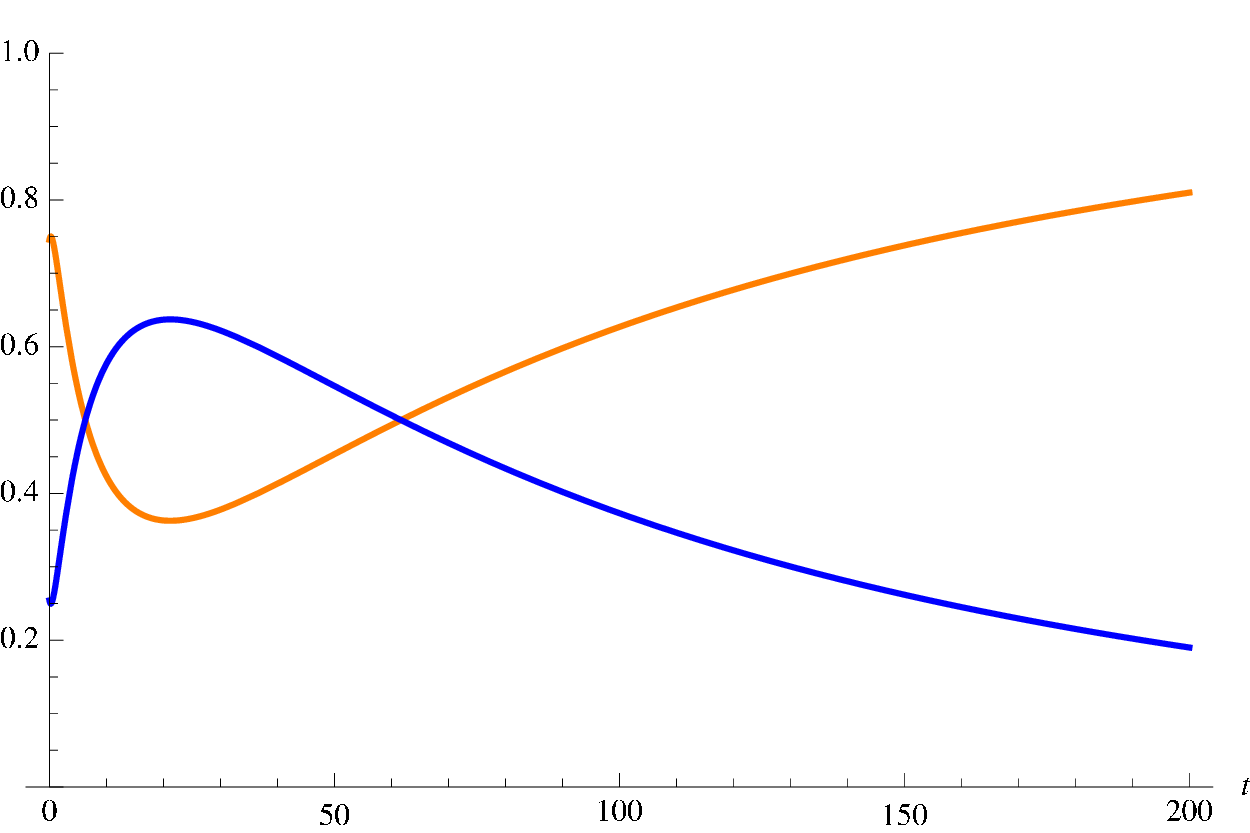}
\end{minipage}
\caption{Time evolution of the modulus square of the overlaps  $a_1(t)$ (orange line) and $a_2(t)$ (blue line) of the wavefunction with the basis vectors for the dipole interaction (left) and for the shell model (right). In these plots $\omega=10$. The initial condition is $c_1$=10, $c_2=0.08$.}\label{fig_1}
\end{figure*}

\begin{figure*}[t]
\begin{minipage}[b]{0.45\linewidth}
\centering
\includegraphics[totalheight=0.2275\textheight]{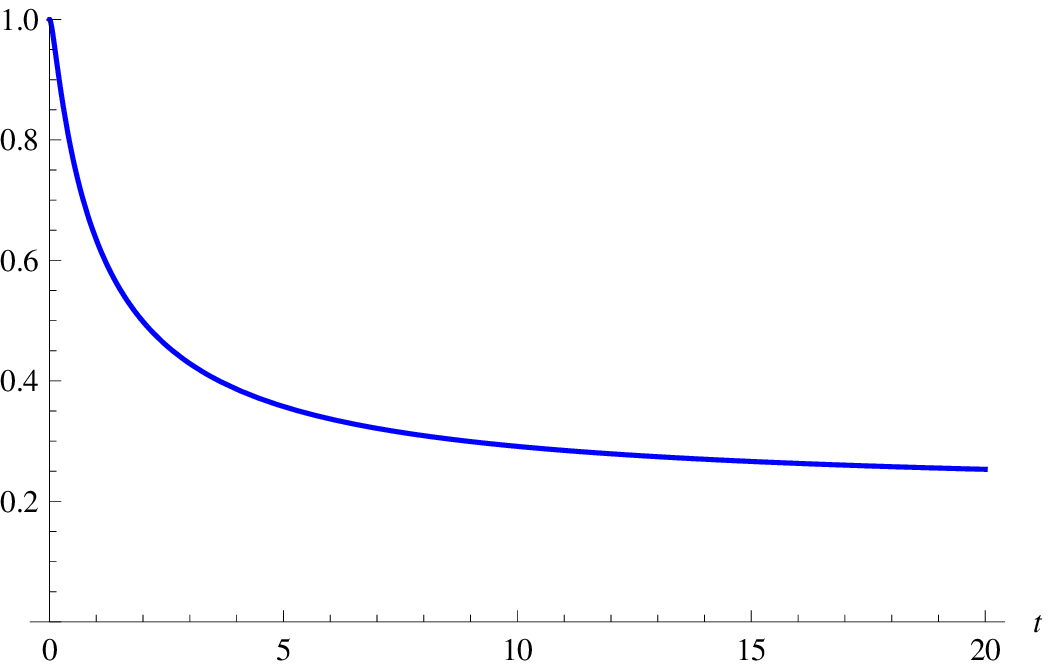}

\end{minipage}
\hspace{0.5cm}
\begin{minipage}[b]{0.45\linewidth}
\centering
\includegraphics[totalheight=0.2275\textheight]{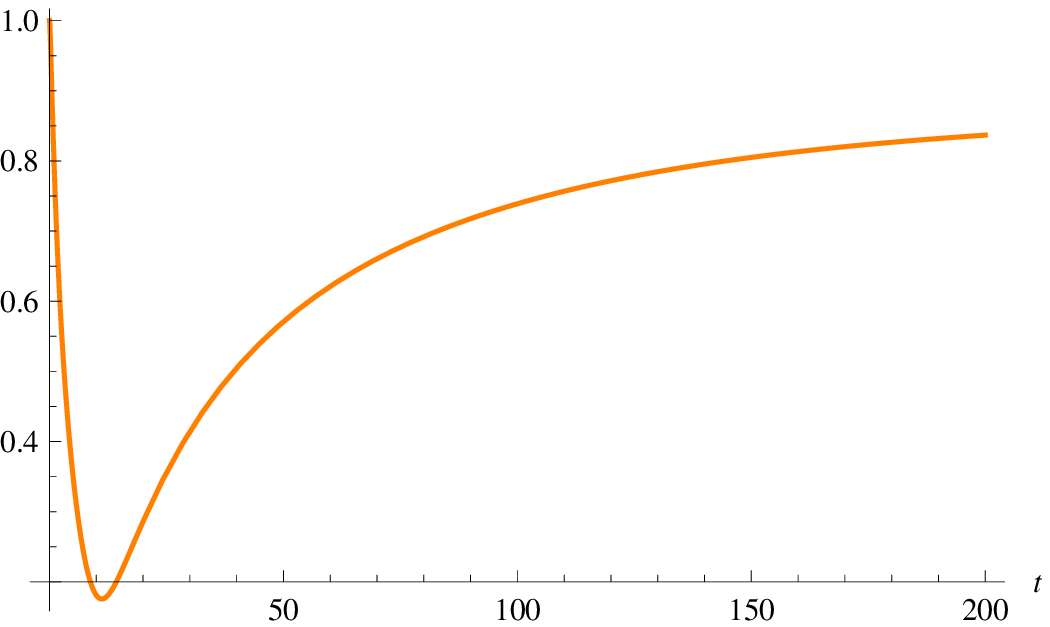}
\end{minipage}
\caption{ Time-dependent fidelity $\left  | \langle \psi(t)| \psi(0)\rangle\right|$ for the dipole interaction (left) and for the shell model (right). As in the previous plots, $\omega=10 $, $c_1=10$ and $c_2=0.08$.}\label{fig_fid}
\end{figure*}

\begin{figure*}[tbh!]
\begin{minipage}[b]{0.45\linewidth}
\centering
\includegraphics[totalheight=0.2275\textheight]{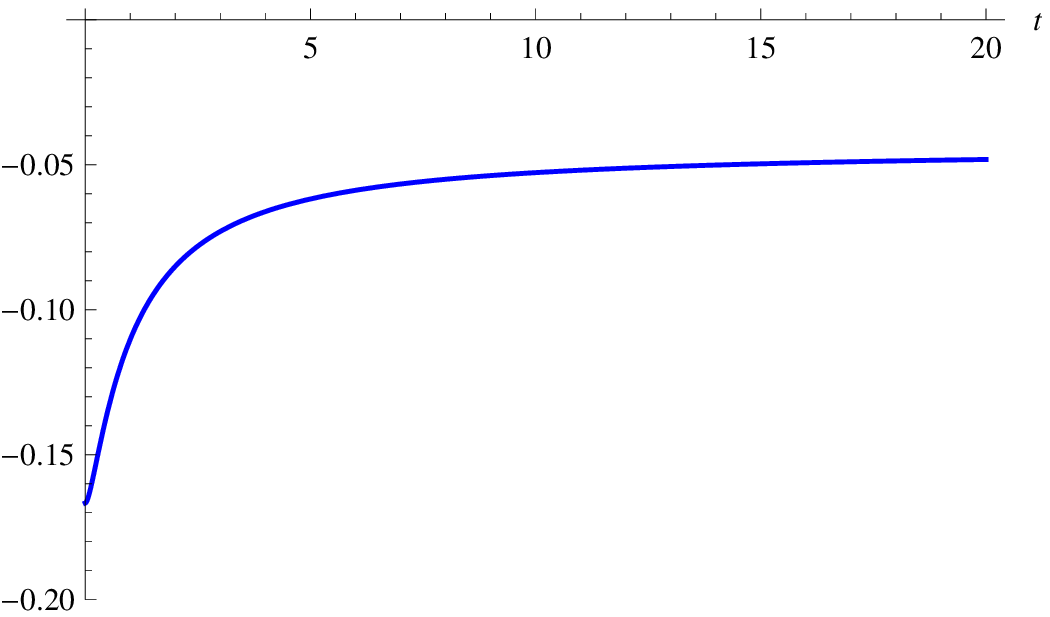}
\includegraphics[totalheight=0.2275\textheight]{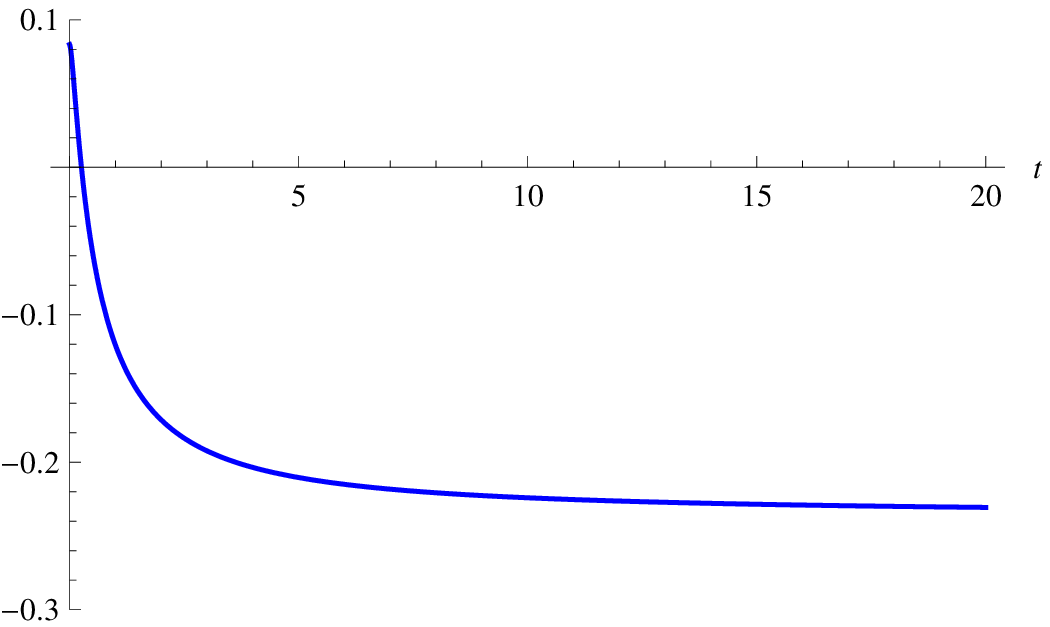}

\end{minipage}
\hspace{0.5cm}
\begin{minipage}[b]{0.45\linewidth}
\centering
\includegraphics[totalheight=0.225\textheight]{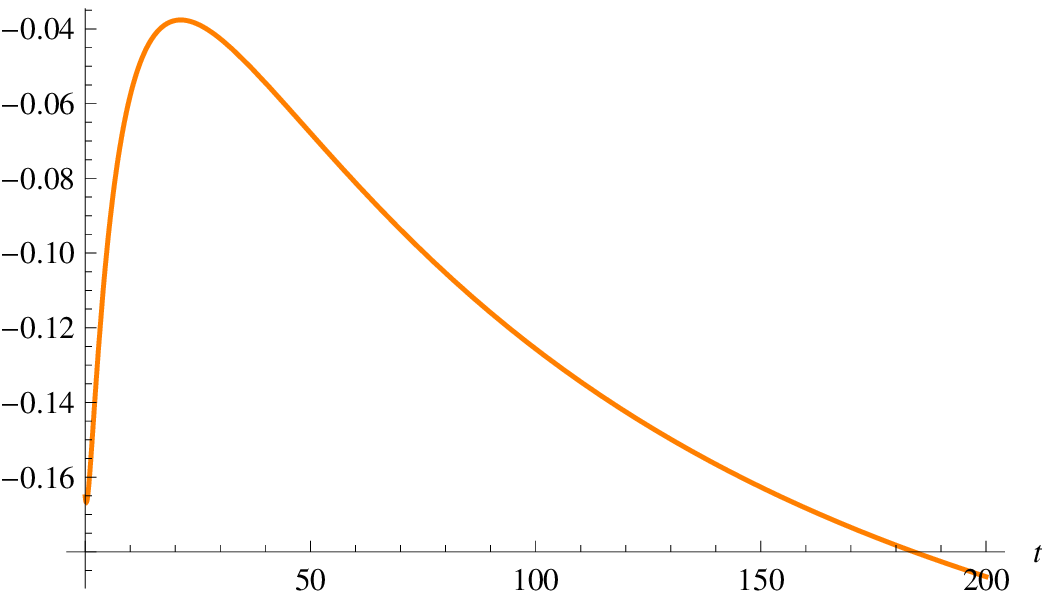}
\includegraphics[totalheight=0.225\textheight]{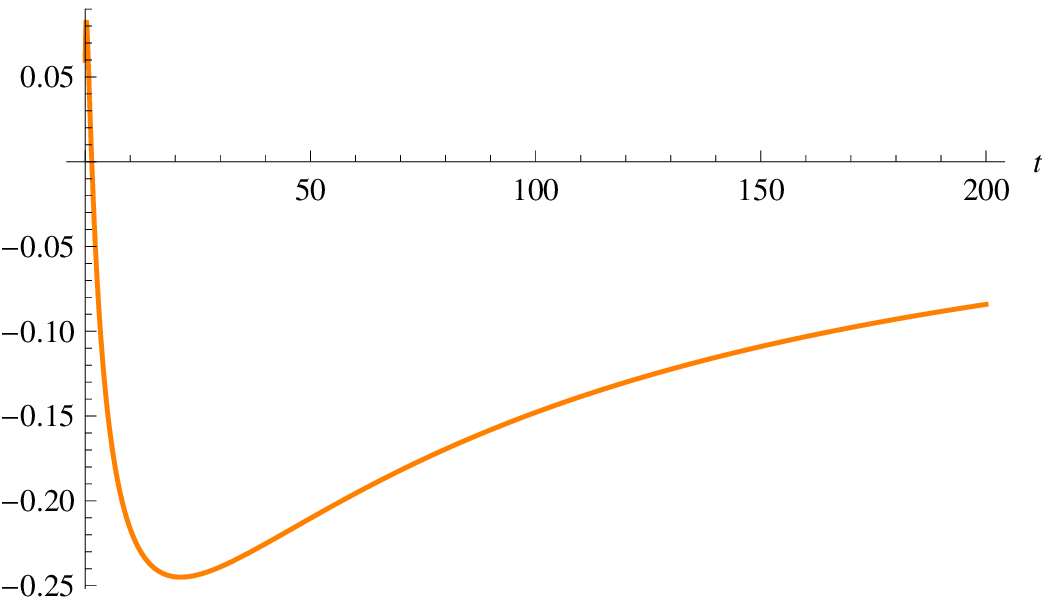}
\end{minipage}
\caption{ Time-dependent correlators $\langle S^z_0(t) S^z_1(t) \rangle$ (top) and $\langle S^z_0(t) S^z_2(t) \rangle$ for the dipole interaction (left) and for the shell model (right). As in the previous plots, $\omega=10$, $c_1=10$ and  $c_2=0.08$.}\label{fig_corr}
\end{figure*}

Two interesting examples are shown in Fig. \ref{fig_1} for the dipolar interaction (left) and for the shell model (right). In both cases, the initial condition is chosen in such a way that it is possible to observe one (dipole interaction) or two (shell model) crossings of the overlaps $a_i(t)$.\\
Another interesting quantity to understand the dynamics of the system is the fidelity $\left  | \langle \psi(t)| \psi(0)\rangle\right|$,  shown in Fig. \ref{fig_fid}, while in  Fig. \ref{fig_corr} we plot the equal times correlators $\langle S^z_0(t) S^z_1(t) \rangle$ and $\langle S^z_0(t) S^z_2(t) \rangle$ ($\langle S^a_j(t)\rangle=0$ for $a=x,y,z$ , as it can be easily understood from (\ref{basis})).

\section{General Solution.}
\label{sec_general}  
It this section, we would like to outline our strategy for getting an integral representation of the solution to the KZ equations (\ref{KZ}) or, more precisely, a generalized version of these equations. Our arguments are a straightforward generalization to the ones of \cite{Babujian,Galitski,Lima-Santos}. The XXZ Gaudin magnets are defined from the Gaudin algebra
\bea
&& \left[S^y (\lambda_1), S^z(\lambda_2) \right]= i \left[Z(\lambda_1-\lambda_2) S^x(\lambda_1)-X(\lambda_1-\lambda_2) S^x(\lambda_2) \right] , \nonumber\\
&& \left[S^x (\lambda_1), S^y(\lambda_2) \right]= i X(\lambda_1-\lambda_2) \left [S^z(\lambda_1)-S^z(\lambda_2) \right] , \nonumber \\
 && \left[S^z (\lambda_1),S^x(\lambda_2) \right]= i \left[X(\lambda_1-\lambda_2) S^y(\lambda_1)-Z(\lambda_1-\lambda_2) S^y(\lambda_2) \right] ,  \label{Gaudin} \\
&& \left[S^a (\lambda_1), S^a(\lambda_2) \right]=0 \nonumber. 
\eea
Here, X and Z are odd functions, with  $Res \left[ X(z)\right]_{z=0}=Res \left[ Z(z)\right]_{z=0}=g$, while the $\lambda_i$'s are complex numbers. Notice that $X$ and $Z$ are not arbitrarily functions, but they have to satisfy a set of quadratic equations that come from the Jacobi  identities for the generators of the Gaudin algebra (\ref{Gaudin}). The solutions to these equations are known, and the simplest one is the rational one $X(\lambda_1,\lambda_2)=Z(\lambda_1-\lambda_2)=\frac{g}{\lambda_1-\lambda_2}$ (for a detailed discussion of Gaudin magnets, the reader is referred to \cite{Ortiz}). For example, in order to describe a spin or fermionic system, the $su(2)$ representation is useful ($S^{\pm}=S^x\pm i S^y$)
\bea
S^{\pm}(\lambda, \underline{z})=\sum_i X(\lambda-z_i) { S^{\pm}_i} ,\\
S^z(\lambda, \underline{z})=-\frac  {\mathds{1}} 2 -  \sum_i Z(\lambda-z_i) S^z_i .
\eea
Here $z_i$ are a set of complex numbers (\emph{the disorder variables}) that are directly linked to the coupling constants of the Hamiltonian, while $S_i$ are the familiar spin operators. Instead, a bosonic system is described by a $su(1,1)$ representation ($\underline{z}=z_0,\ldots z_{N-1}$)
\bea
S^{\pm}(\lambda, \underline{z})=\pm\sum_i X(\lambda-z_i) { K^{\pm}_i} ,\\
S^z(\lambda, \underline{z})=-\frac {\mathds{1}} 2 - \sum_i Z(\lambda-z_i) K^z_i .
\eea
where the $K_i$'s satisfy a $su(1,1)$ algebra. Of course, mixed representations are also available, in order to describe a system where the spin degrees of freedom interact with a bosonic bath ({\it i.e.} the Dicke model and its generalizations). In the following, we will use the notation $A^a_i=S^a_i / K^a_i$ respectively for the $su(2)$ representation and the $su(1,1)$ one.

From the Gaudin algebra, it is possible to define the generating function
\be
H(\omega,\underline{z})=\sum_{a=x,y,z} S^a(\omega,\underline{z}) S^a(\omega,\underline{z}) ,
\ee
and the integrals  of motion
\bea
H_i(\underline{z})=\frac 1 {2 g^2}Res \left[H(\omega,\underline{z})\right]_{\omega=z_i}.
\eea
These $H_i$'s are a family of commuting operators, and each of them can be considered as the Hamiltonian of a quantum system. Since  $H(w,\underline{z})$ can be diagonalized using the algebraic Bethe Ansatz, the Hamitonians $H_i$ are exactly solvable. In the rational case, the integrals of motion for a spin system reduce to
\be
H_i(\underline{z})=\frac 1 {2 g} S^z_i+\sum_{j\ne i} \frac{\mathbf{S}_i \cdot \mathbf{S}_j}{z_i-z_j} ,
\ee
that is a central spin Hamiltonian with a Zeeman term for the central spin: when the magnetic field is zero it reduces to the operators appearing in the KZ equations (\ref{KZ}). The main statement of this section is that, for the spin as well as for the bosonic representation, it is possible to construct an explicit solution of the generalized KZ equations
\be
k \frac {\partial \Psi_N} {\partial z_i} (\underline{z})=H_i(\underline{z}) \Psi_N(\underline{z}) . \label{gen_KZ}
\ee
This equation, as we have explained in the first section of this paper, is not simply a mathematical curiosity, but it can be linked to a time dependent Schr{\"o}dinger equation: $\psi_N(t)=\Psi_N(z_0(t),\ldots,z_{N-1}(t))$  satisfy the Schr{\"o}dinger equation with Hamiltonian $H(t)=\frac {\hbar\,i} k \sum_j\dot{z}_j(t) H_j(\underline{z}(t))$. Quite nicely, this integral representation of the solution is based on the integrability of the model. Let us introduce the Bethe state ($\underline{\lambda}=\lambda_1,\ldots, \lambda_M$ )
\be
|\Phi(\underline{\lambda},\underline{z})\rangle=\prod_{\alpha=1}^M S^+(\lambda_\alpha, \underline{z}) |0\rangle,
\ee
where the vacuum $|0\rangle$ is annihilated by $S^-(\lambda)$ and is an eigenstate of $S^z(\lambda)$ ($A_i^z|0\rangle=d_i |0\rangle$). It can be proven that this state obeys the off-shell Bethe equation
\bea
&&H_i(\underline{z}) |\Phi(\underline{\lambda},\underline{z})\rangle=h_i(\underline{\lambda},\underline{z})|\Phi(\underline{\lambda},\underline{z})\rangle+\\
&&+\sum_{\alpha} X(\lambda_\alpha-z_i) f_\alpha(\underline{\lambda},\underline{z}) A^+_j\prod_{\beta \ne \alpha}S^+(\lambda_\beta,\underline{z}) |0\rangle\nonumber.
\eea
The functions $h_i$ and $f^\alpha$ can be derived from a Yang-Yang action: more precisely, $h_i=\frac {\de {\cal S}}{\de z_i}$, $f_\alpha= \frac {\de {\cal S}}{\de \lambda_\alpha}$, where ${\cal S}$ is defined as
\bea
{\cal S}(\underline{\lambda},\underline{z})=&&\sum_{i}\frac{z_i d_i}{2 g}+\sum_{i}\sum_{j\ne i} \frac{d_i d_j}{2g} T(z_i-z_j)+\sum_i \sum_\alpha^M d_i T(z_i-\lambda_\alpha)+ \nonumber \\
&&+\sum_\alpha^M \frac{\lambda_\alpha}{2 g}+\sum_\alpha^M\sum_{\beta \ne \alpha}^M \frac {T(\lambda_\beta-\lambda_\alpha)}{2 g} ,\label{action}
\eea
where $T(u)=\int^u dz Z(z)$.
Usually, one imposes the on-shell condition $f^\alpha=0$ (Bethe equations), thus obtaining a basis of eigenstates of the Hamiltonian. Here instead, we take advantage of the existence of the action $\cal S$ and we define

 \begin{equation}
\Psi_N(\underline{z})=\oint_\gamma d \underline{\lambda} \, \,e^{\frac{{\cal S} (\underline{\lambda},\underline{z})}k }  |\Phi(\underline{\lambda},\underline{z})\rangle, \label{sol}
\end{equation}
where the closed contour $\gamma$ is chosen in such a way that the branch of the integrand at the end point of $\gamma$ is  the same as that at the initial point. It is quite easy to show that  (\ref{sol}) is indeed a solution of (\ref{gen_KZ}). Notice that  due to the multi-valuedness of  the integrand, the path of integration is usually highly nontrivial, this being the major technical difficulty of our approach. In the rational case, these integrals represent multivariable hypergeometric functions \cite{Aomoto,love}, and our hope is that this connection could be exploited to evaluate explicitly \ref{sol}.

\subsection{\it The $\it{k}\to0$ limit and the completeness of the integral representation} \label{sub_adiabatic}
Unfortunately, a direct evaluation of (\ref{sol}) is beyond our present ability. However, in the $k\to 0$ limit, the only contribution to the integral comes from the stationary points of the action (\ref{action}), i.e. from the on-shell Bethe state \cite{Reshetikhin}. Therefore, it is quite interesting to discuss the physical meaning of this limit for our time-dependent Schr{\"o}dinger equation. The most natural interpretation of this limit is as an adiabatic one. As an example, let us consider the central spin limit with coupling constant (\ref{coupling}). If we parametrize $k=i \,v$ and $z_0(t)=F(\Omega_0 \,v\, t)$, where $F$ is an arbitrary function, we have a central spin model with coupling constants
 \be
 J_i(t)=\hbar\, \Omega_0 \frac { F'(\Omega_0 \,v \,t)} {F(\Omega_0\, v\, t)-z_i}.
 \ee
 The time scale of $J_i(t)$ is $\Omega_0 v$, and so when $v\to0$ the system is in an adiabatic regime. Notice that, indeed, the contribution from the stationary points of (\ref{sol}) agrees completely with the usual quantum adiabatic theorem: the stationary condition $f_\alpha=\frac {\de {\cal S}}{\de \lambda_\alpha}=0$ imposes the Bethe equations, thus selecting the instantaneous eigenstate of the Hamiltonian, while $\exp\left[ \frac{{\cal S} (\underline{\lambda},\underline{z})}k \right]_{f_\alpha=0}$ is the corresponding dynamical phase. Moreover, by choosing properly the contour $\gamma$, we can select any eigenstate, and therefore our solution is complete in the adiabatic limit. This is at least a strong hint that our solution is complete for any time dependency of the coupling constants.\\
 Quite interestingly, the $k\to 0$ limit can be interpreted also as a semiclassical limit, if we use a different parametrization. Indeed, if $ k= i \frac{\hbar}{{\cal A}_0}$, where ${\cal A}_0$ has the dimensions of an action, we have a central spin model with coupling constants
\be
 J_{i}(t)= {\cal A}_0 \frac{\dot{z}_0(t)}{z_0(t)-z_i},
 \ee
 and $k\to0$ is equivalent to ${\cal A}_0\gg \hbar$.\\
 
 \section{Conclusions.}
 \label{sec_conclusions}
In this article, we have studied a class of time-dependent Hamiltonians which possess many-body wavefunctions given by solutions to the associated Knizhnik-Zamolodchikov equations. The underlying time-independent integrability and link with CFT allow us to provide an explicit integral representation of the solution to the time-dependent Schr{\"o}dinger equation. For a small system, these solutions reduce to the  familiar hypergeometric functions, allowing us to easily study the dynamics of the system, as we did for a 4 spins model. This specific example shows that the exact solubility of these time dependent systems is not due to their triviality. Instead, their physics appears to be quite rich, as you could expect for a full time dependent problem.\\
Of course, from a practical point of view, the most interesting thing would be to solve a N-particle model. In order to do so, we have to deal with the complicated integral (\ref{sol}). Unfortunately, we are not able to do it at the present time. However, in the rational case, this integral reduces to generalized hypergeometric functions, that have been extensively studied in the mathematical literature. Another possible line of investigation could be to compute the corrections to the adiabatic limit, that could teach us something about this complicated integral representation.\\
While our construction works only if the time dependence of the Hamiltonian is finely tuned, it provides an intriguing starting point for understanding the consequences of quantum integrability in time-dependent physics.\\

DF and VG are supported by the Swiss NSF under grants PP00P2\_140826 and NSF PHY11-25915. J-S C and VG thank KITP for hospitality.  J-S C acknowledges support from the Foundation for Fundamental Research on Matter (FOM) and from the Netherlands Organization for Scientific Research (NWO).

\bibliographystyle{vancouver}

\clearpage

\appendix
\section{The Central Spin Model and the WZW CFT}
\label{app_wzw}
In the text, we have argued that the conformal block of the $SU(2)_k$ $\Psi_N(z_0,\ldots,z_{N-1})$ can be used to construct the wave function $\psi_N(t)=\Psi_N(z_0(t),z_1,\ldots,z_{N-1})$, that is a solution of the Schr{\" o}dinger equation for the time dependent central spin Hamiltonian ($k=i\,v$)
\be
H_{CS}(t)=\sum_{i=1}^{N-1}  \frac{\hbar \,\dot{z}_0(t)}{v\,(z_0(t)-z_i(t))}   {\bf S}_{0}\cdot {\bf S}_{i}.
\ee
This result follows easily from the K-Z equations satisfied by the conformal blocks
\be
\left[k \frac {\de} {\de z_i}-\sum_{j\ne i} \frac{\mathbf{S}_i \cdot \mathbf{S}_j}{z_i-z_j}\right] \Psi_N(z_0,\ldots,z_{N-1})=0. 
\ee
Here, we would like to elaborate more on this point. In particular, one could wonder if there exists a more general wavefunction $\psi_N(t)=\Psi_N(z_0(t),z_1(t),\ldots,z_{N-1}(t))$, where all the $z_i(t)$ are time dependent, that satisfies the Schr{\" o}dinger equation for a central spin model. The answer is essentially no. Indeed, $\psi_N(t)=\Psi_N(z_0(t),z_1(t),\ldots,z_{N-1}(t))$, satisfies a Schr{\" o}dinger equation with Hamiltonian
\bea
&&H(t)=\sum_i \sum_{j < i} C_{ij} (t) {\mathbf{S}_i \cdot \mathbf{S}_j} ,\\
&& C_{i j}(t)=\frac{\hbar\,\left[\dot{z}_i(t)-\dot{z}_j(t) \right]}{v\left[z_i(t)-z_j(t) \right]}.
\eea
So, if we want to cancel out the couplings between the spins of the bath, we need to have $\dot{z}_1(t)=\dot{z}_2(t)\ldots=\dot{z}_{N-1}(t)$.This is equivalent to have the point $z_0(t)$ moving in time while $z_1,\ldots,z_{N-1}$ are fixed, once we take into account the invariance of the conformal block $\Psi_N(z_0,z_1,\ldots,z_{N-1})$ under global translations.
\section{The Four Spins Conformal Block}
\label{app_four}
In this section,  we would like to derive explicitly the four point conformal block $\Psi_4(z_0,z_1,z_2,z_3)$. The ${\bf S}^2=0$ subspace is spanned by two vectors
\bea
&&\!\!\!\!\!  \!\!\!\!\!  \!\!\!\!\!  \!\!\!\!\!  \!\!\!\!\! |v_1\rangle=\left(\frac{|+-\rangle-|-+\rangle}{\sqrt{2}}\right) \otimes \left(\frac{ |+-\rangle-|-+\rangle }{\sqrt{2}}\right)\\
&& \!\!\!\!\!  \!\!\!\!\!  \!\!\!\!\!  \!\!\!\!\! \!\!\!\!\! |v_2\rangle=\frac 1 {\sqrt{3}}\left[ |++--\rangle+ |--++\rangle +\left(\frac{ |+-\rangle+|-+\rangle }{\sqrt{2}}\right)\otimes \left(\frac{ |+-\rangle+|-+\rangle }{\sqrt{2}}\right) \right] \nonumber .
\eea
It turns out that 
\bea
&& {\bf S}_0\cdot {\bf S}_1 |v_1 \rangle=-\frac {3}{4} |v_1\rangle \qquad {\bf S}_0\cdot {\bf S}_1 |v_2 \rangle=\frac {1}{4} |v_2\rangle\\
&& {\bf S}_0\cdot {\bf S}_2 |v_1 \rangle =-\frac{\sqrt{3}}{4} |v_2\rangle \qquad {\bf S}_0\cdot {\bf S}_2 |v_2 \rangle =-\frac{\sqrt{3}}{4} |v_1\rangle-\frac 1 2 |v_2 \rangle .
\eea
Moreover, in this subspace $S^a|\psi\rangle =\sum_j S^a_j |\psi\rangle=0$, hence
\be
\frac 3 4 + {\bf S}_0\cdot {\bf S}_1+ {\bf S}_0\cdot {\bf S}_2+ {\bf S}_0\cdot {\bf S}_3  =0 \,\,\,({\bf S}^2=0) . \label{sum_rule}
\ee
So, let us consider the $z_0$ K-Z equation 
\be
\left [ k \frac \de {\de z_0}-\frac{{\bf S}_0\cdot {\bf S}_1 }{z_0-z_1} -\frac{{\bf S}_0\cdot {\bf S}_2 }{z_0-z_2}- \frac{{\bf S}_0\cdot {\bf S}_3}{z_0-z_3} \right] \Psi_4(z_0,z_1,z_2,z_3)=0.
\ee
The sum rule (\ref{sum_rule}) teaches us that we can eliminate the dependence of the Hamiltonian on $ {\bf S}_0\cdot {\bf S}_3$.Moreover, if we make the substitution
\be
 \Psi_4(z_0,z_1,z_2,z_3)=\left[(z_0-z_3)(z_1-z_2)\right]^{\frac{i 3} 4} |\varphi(z_0,z_1,z_2,z_3) \rangle, \label{phase}
\ee
the  K-Z equation become
\be
\!\!\!\!\!  \!\!\!\!\!  \!\!\!\!\!  \!\!\!\!\!  \!\!\!\!\!  \!\!\!\!\! \left[i \frac \de {\de z_0}+
{\bf S}_0\cdot {\bf S}_1 
 \left(\frac 1{z_0-z_3}-\frac 1{z_0-z_1}\right)+{\bf S}_0\cdot {\bf S}_2 \left(\frac 1{z_0-z_3}-\frac 1{z_0-z_2}\right) \right] |\varphi \rangle=0.
\ee
Let us now introduce the new variable 
\be
x=\frac{(z_0-z_1)(z_2-z_3)}{(z_0-z_3)(z_2-z_1)},
\ee
so
\be
\frac \de  {\de z_0}=x\left[ \frac 1{z_0-z_1}-\frac 1{z_0-z_3}\right] \frac \de  {\de x}.
\ee
Moreover, the following identity holds:
\be
\frac x {1-x}\left[ \frac 1{z_0-z_1}-\frac 1{z_0-z_3}\right]=\left[ \frac 1{z_0-z_3}-\frac 1{z_0-z_2}\right].
\ee
This identity can be proved by a brute force calculation (it's just a trivial algebraic calculation), but it is possible to obtain it in a clever way. First of all,we notice that the l.h.s. has poles in $z_0=z_3$ and $z_0=z_2$ (since $x=1$), so we can deduce that
\be
l.h.s.=\frac {Q(z_0,z_1,z_2,z_3)}{(z_0-z_2)  (z_0-z_3)},
\ee
where $Q$ is a polynomial. However, the l.h.s is zero only for $z_2=z_3$, hence $Q=A(z_3-z_2)$. Taking the limit $z_0\to z_3$, we get $A=1$.\\
Therefore, the K-Z equation reduces to
\be
\left[i \frac \de {\de x}-\frac{{\bf S}_0\cdot {\bf S}_1 } x-\frac{{\bf S}_0\cdot {\bf S}_2}{x-1} \right] |\varphi(x,z_1,z_2,z_3) \rangle=0.
\ee
Now, we can expand $|\varphi \rangle$ on our basis
\be
|\varphi(x,z_1,z_2,z_3) \rangle=\sum_{i=1}^2 F_i(x,z_1,z_2,z_3) |v_i\rangle, \label{vector}
\ee
obtaining a system of differential equations for $F_1$ and $F_2$
\bea
&&i  F'_1 + \frac 3 4 \frac{F_1}{x}+\frac{\sqrt{3}}{4}\frac{F_2}{x-1}=0\\
&& i  F'_2- \frac 1 4 \frac{F_2}{x}+\frac 1 {x-1} \left[\frac{\sqrt{3}} 4 F_1+\frac 1 2 F_2 \right],
\eea
where the prime denote a derivative respect to x.
Our aim now is to show that these equations reduce to a hypergeometric one. From the first one we get
\be
F_2=-(x-1) \left[ \frac{\sqrt{3}F_1}{x}+\frac{4 i}{\sqrt{3}} F'_1\right],
\ee
and, the second equation reduces, after some algebra, to
\be
\alpha(x)F_1''(x)+\beta(x) F_1'(x)+\gamma(x) F_1(x)=0,
\ee
where
\bea
&&\alpha(x)=-\frac 4 {\sqrt{3}} (1-x)\\
&&\beta(x)=\frac {2\, i}{x \sqrt{3}} +\frac 4{\sqrt{3}} (1-i)\\
&&\gamma(x)=\sqrt{3}\left\{- \frac 1{x^2}\left[i+\frac 1 4 \right] -\frac 1 4 \frac 1 x + \frac 1 4 \frac 1 {x-1}\right\} ,
\eea
or, equivalently, 
\be
\!\!\!\!\!  \!\!\!\!\!  \!\!\!\!\! \!\!\!\!\! \!\!\!\!\! 
x (1-x) F_1''(x)+ \left[ \frac i 2 + (1-i)x\right]F_1'(x)-\frac 3 {16} \left[ \frac {4\, i +1}{x}+1 -\frac 1 {x-1} \right] F_1(x)=0 . \label{eq}
\ee
This equation is indeed quite similar to the hypergeometric equation
\be
x (1-x) w''(x)+\left[ c- (a+b+1) x \right] w'(x)-ab\,\,w(x)=0,
\ee
but not identical, since the coefficient of the last term depends on x. How can we get rid of those terms? Let us define a function $v(x)$ such that
\be
F_1(x)=x^r v(x),
\ee
where $r$ is a complex number. Clearly,
\bea
&&F'_1(x)=r x^{r-1} v(x)+x^r v'(x), \\
&&F''_1(x)=r (r-1) x^{r-2} v(x)+2r x^{r-1} v'(x)+ x^r v''(x),
\eea
so, if we substitute these expressions in (\ref{eq}) we get that the term proportional to $v(x)$ is
\be
\frac 1 x \left[ r (r-1)+\frac i 2 r-\frac{3(4\,i+1)}{16} \right]+\cdots
\ee
where we omitted terms regular for $x=0$. So, if we choose $r=\frac{3\,i} {4}$ or $r=1-\frac i 4$, we can eliminate the nasty $\frac 1 x$ term. A similar argument applies for $x-1$. So, let us define a function $g(x)$ such that
\be
F_1(x)=x^{\frac {3 i}{4} }(x-1)^{-\frac i 4}\, g(x).
\ee
It turns out that $g(x)$ satisfy a hypergeometric equation
\be
x (1-x) g''(x)+ (i-x) g'(x)-\frac 1 4 g(x)=0,
\ee
with parameters $a=-b=\frac i 2$, $c=i$.So, if $w_1$ and $w_2$ are two linearly independent solutions of the hypergeometric equation, we have 
\be
F_1(x)=x^{\frac {3 i}{4} }(x-1)^{-\frac i 4} \left[ c_1 w_1(x) + c_2 w_2(x)\right]
\ee
where $c_1$ and $c_2$ are arbitrarily constant, determined by the initial condition.
\subsection{A simple time evolution.}
Let us discuss now a simple example. At time t=0, the spins $\mathbf{S}_i$, $i=1,2,3$, are at a distance j from the central spin $\mathbf{S}_0$. Therefore, their coupling with the central spin $J_i$  is proportional to $\exp\left(-i^2\right)$. Then, for $t>0$, the coupling constant decreases linearly in time. So, we can model this situation if $z_0=\omega t$, and $z_i= \exp\left(i^2\right)$ $i=1,2,3$. Therefore,
\be
H(t)=\omega\left[ \frac {{\bf S}_0\cdot {\bf S}_1 }{\omega \, t+e}+\frac {{\bf S}_0\cdot {\bf S}_2}{\omega \, t+e^4}+\frac {{\bf S}_0\cdot {\bf S}_3}{\omega \, t+e^9} \right],
\ee
while 
\be
x(t)=- \left[\frac{e^9-e^4}{e^4-e} \right]\, \left[\frac{\omega t + e} {\omega t + e^9}  \right]      \in \left(-\infty,0\right).
\ee
So, there are 24 solutions of the hypergeometric equation in the complex plane~\cite{love}. These solutions are characterized by different analytical properties: for example the hypergeometric function $F(a,b,c,z)$ is  analytic in 0, while $F(a,b,a+b+1-c,1-x)$ is analytic in 1. Of course, if different solutions are well defined in the same region, at most two of them can be linearly independent, since the hypergeometric equation is of second order. For our aims, a good choice of two
 linearly independent solutions is  $F(a,b,c,z)$ and $z^{1-c}F(1+a-c,1+b-c,2-c,z)$, since they have no singularity on the negative real line. So, we put
\bea
&&w_1(x)=F\left(\frac i 2,-\frac i 2,i,x\right)\\
&&w_2(x)=x^{1-i} F\left(1-\frac{ i} 2,1-\frac{3\, i} 2,2-i,x\right)
\eea
while $F_2(x)=c_1 y_1(x)+c_2 y_2(x)$ is thus given by
\bea
&&\!\!\!\!\! \!\!\!\!\! \!\!\!\!\! \!\!\!\!\! \!\!\!\!\! \!\!\!\!\! y_1(x)=\frac{(x-1)^{- \frac i 4} x^{\frac {3 i}{ 4}}} {\sqrt{3}}\left\{ - x^i \left[F\left(-\frac i 2,\frac i 2, i, x\right)+(x-1) F\left(1-\frac i 2,1+ \frac i 2,1+i,x\right) \right]\right\}  \nonumber ,\\
&&\!\!\!\!\! \!\!\!\!\! \!\!\!\!\! \!\!\!\!\! \!\!\!\!\! \!\!\!\!\!  y_2(x)=\frac{(x-1)^{- \frac i 4} x^{\frac {3i}{ 4}} }{\sqrt{3}} \Bigg\{ \left[ 2-3x\right] F\left( 1- \frac i 2,1-\frac{3\, i} 2,2-i,x\right)+\\ 
&&\!\!\!\!\! \!\!\!\!\! \!\!\!\!\! \!\!\!\!\! \!\!\!\!\! \!\!\!\!\! -\left[ 2+4\, i\right] \left[x-1 \right] F \left(1-\frac{3 \, i} 2,2-\frac i 2,2-i,x \right)\Bigg\} \nonumber,
\eea
and so
\be
F_i(x)=M_{i j}(x) c_j,\qquad M(x)=
\left[ 
\begin{array}{c c}
w_1(x) & w_2(x)\\
 y_1(x) & y_2(x)
\end{array}
\right]
.
\ee

As we have discussed in the text, the physics of this simple model could be quite interesting, with a double crossing of the overlaps $|F_i|^2$ for some initial conditions. Moreover, as it is shown in Fig. 4  , the absolute value of the determinant of M(x) is constant for $x<0$ and non vanishing. This indeed imply that this class of solution spans the whole ${\bf S}^2=0$ subspace.

\begin{figure*}[t]
\begin{minipage}[b]{0.45\linewidth}
\centering
\includegraphics[totalheight=0.2\textheight]{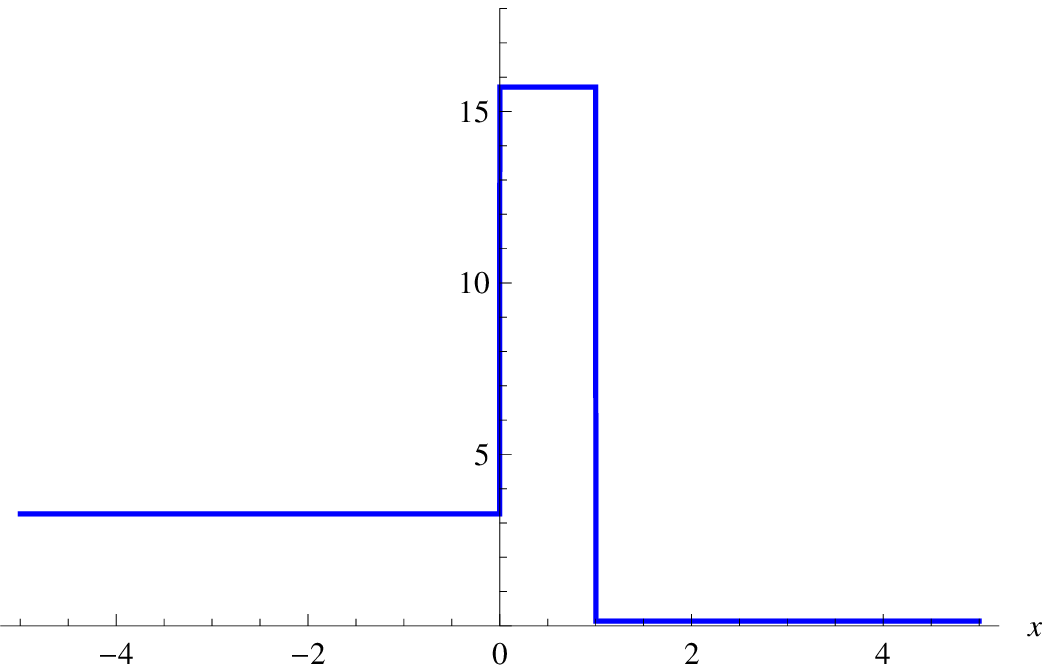}
\end{minipage}
\hspace{0.3cm}
\begin{minipage}[b]{0.45\linewidth}
\centering
\includegraphics[totalheight=0.2\textheight]{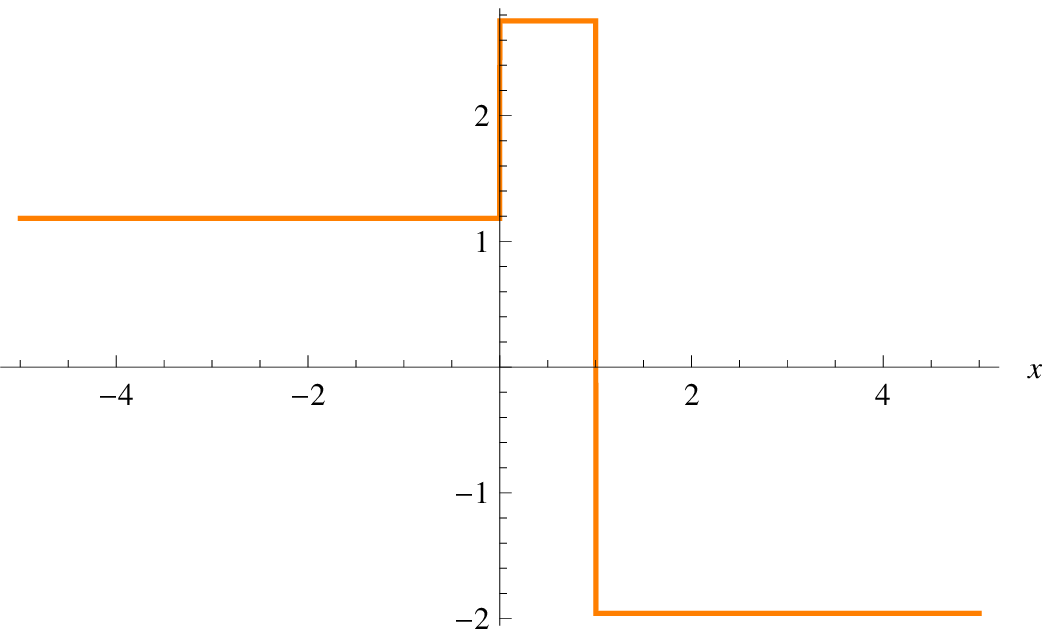}
\end{minipage}
\caption{$|\det[M(x)]|$ (left) and $\log\left[|\det[M(x)]| \right]$ (right) as functions of x.}\label{fig_det}
\end{figure*}

More precisely,  $|\det[M(x)]|$ is indeed a (non vanishing) constant in the connected domains $( -\infty, 0)$ , $(0,1)$ and $(1, +\infty)$, with jumps at the singular points of $w_1$ and $w_2$.  The fact that, inside these domains, 
 $|\det[M(x)]|$  is a non zero constant reflects the unitarity of the quantum time evolution, that implies 
 \be
|F_1(x)|^2+|F_2(x)|^2=\textrm{non vanishing constant},
\ee
or, equivalently
\be
M^\dagger(x) M(x)=\textrm{non vanishing constant},
\ee
and therefore
\be
|\det(M(x)|^2=\textrm{non vanishing constant}.
\ee

\bibliography{kzpaper}

\begin{thebibliography}{10}

\bibitem{quench_Calabrese2006}
Calabrese P, Cardy J.
\newblock Time Dependence of Correlation Functions Following a Quantum Quench.
\newblock Physical Review Letters. 2006;96(136801).

\bibitem{RMP}
Polkovnikov A, Sengupta K, Silva A, Vengalattore M.
\newblock \textit{Colloquium} :Nonequilibrium dynamics of closed interacting
  quantum systems.
\newblock Reviews of Modern Physics. 2011;83(3):863--883.

\bibitem{ent_Calabrese}
Calabrese P, Cardy J.
\newblock Evolution of entanglement entropy in one-dimensional systems.
\newblock Journal of Statistical Mechanics: Theory and Experiment.
  2005;2005(04):P04010.
\newblock Available from:
  \url{http://stacks.iop.org/1742-5468/2005/i=04/a=P04010}.

\bibitem{ent_DeChiara}
Chiara GD, Montangero S, Calabrese P, Fazio R.
\newblock Entanglement entropy dynamics of Heisenberg chains.
\newblock Journal of Statistical Mechanics: Theory and Experiment.
  2006;2006(03):P03001.
\newblock Available from:
  \url{http://stacks.iop.org/1742-5468/2006/i=03/a=P03001}.

\bibitem{ent_Lauchli}
L{\"a}uchli AM, Kollath C.
\newblock Spreading of correlations and entanglement after a quench in the
  one-dimensional Bose--Hubbard model.
\newblock Journal of Statistical Mechanics: Theory and Experiment.
  2008;2008(05):P05018.
\newblock Available from:
  \url{http://stacks.iop.org/1742-5468/2008/i=05/a=P05018}.

\bibitem{ent_Alba}
Alba V, Heidrich-Meisner F. 2014 02;Available from:
  \url{http://arxiv.org/abs/1402.2299}.

\bibitem{Lewis}
Lewis HRJ, Riesenfeld WB.
\newblock An Exact Quantum Theory of the Time‐Dependent Harmonic Oscillator
  and of a Charged Particle in a Time-Dependent Electromagnetic Field.
\newblock Journal of Mathematical Physics. 1969;10:1458.

\bibitem{Perelomov}
Perelomov AM, Popov VS.
\newblock Group-theoretical aspects of the variable frequency oscillator
  problem.
\newblock Theoretical and Mathematical Physics. 1969;1(3):275--285.

\bibitem{Kagan}
Kagan Y, Surkov EL, Shlyapnikov GV.
\newblock Evolution of a Bose-condensed gas under variations of the confining
  potential.
\newblock Physical Review A. 1996;54(3):R1753--R1756.

\bibitem{Pitaevskii}
Pitaevskii LP, Rosch A.
\newblock Breathing modes and hidden symmetry of trapped atoms in two
  dimensions.
\newblock Physical Review A. 1997;55(2):R853--R856.

\bibitem{Minguzzi}
Minguzzi A, Gangardt DM.
\newblock Exact Coherent States of a Harmonically Confined Tonks-Girardeau Gas.
\newblock Physical Review Letters. 2005;94:240404.

\bibitem{Gritsev}
Gritsev V, Barmettler P, Demler E.
\newblock Scaling approach to quantum non-equilibrium dynamics of many-body
  systems.
\newblock New Journal of Physics. 2010;12(11):113005.

\bibitem{Moore}
Moore G, Read N.
\newblock Nonabelions in the fractional quantum hall effect.
\newblock Nuclear Physics B. 1991;360(2-3):362 -- 396.

\bibitem{Sierra}
Sierra G.
\newblock Conformal field theory and the exact solution of the BCS Hamiltonian.
\newblock Nuclear Physics B. 2000;572(3):517 -- 534.

\bibitem{Galitski}
Sedrakyan TA, Galitski V.
\newblock Boundary Wess-Zumino-Novikov-Witten model from the pairing
  Hamiltonian.
\newblock Physical Review B. 2010;82(21).

\bibitem{GaudinBOOK}
Gaudin M.
\newblock La fonction d'onde de Bethe.
\newblock Masson; 1983.

\bibitem{Amico}
Amico L, Falci G, Fazio R.
\newblock The BCS model and the off-shell Bethe ansatz for vertex models.
\newblock Journal of Physics A: Mathematical and General. 2001;34(33):6425.

\bibitem{Ortiz}
Ortiz G, Somma R, Dukelsky J, Rombouts S.
\newblock Exactly-solvable models derived from a generalized Gaudin algebra.
\newblock Nuclear Physics B. 2005;707(3):421 -- 457.

\bibitem{ReviewQD}
Urbaszek B, Marie X, Amand T, Krebs O, Voisin P, Maletinsky P, et~al.
\newblock Nuclear spin physics in quantum dots: An optical investigation.
\newblock Reviews of Modern Physics. 2013;85(1):79--133.

\bibitem{Imamoglu}
Latta C, Srivastava A, Imamoglu A.
\newblock Hyperfine Interaction-Dominated Dynamics of Nuclear Spins in
  Self-Assembled InGaAs Quantum Dots.
\newblock Physical Review Letters. 2011;107(16).

\bibitem{Lukin}
Childress L, Gurudev~Dutt MV, Taylor JM, Zibrov AS, Jelezko F, Wrachtrup J,
  et~al.
\newblock Coherent Dynamics of Coupled Electron and Nuclear Spin Qubits in
  Diamond.
\newblock Science. 2006;314(5797):281--285.

\bibitem{DFS1}
Lidar DA, Chuang IL, Whaley KB.
\newblock Decoherence-Free Subspaces for Quantum Computation.
\newblock Physical Review Letters. 1998;81:2594--2597.
\newblock Available from:
  \url{http://link.aps.org/doi/10.1103/PhysRevLett.81.2594}.

\bibitem{DFS2}
Zanardi P, Rasetti M.
\newblock Noiseless Quantum Codes.
\newblock Physical Review Letters. 1997;79:3306--3309.
\newblock Available from:
  \url{http://link.aps.org/doi/10.1103/PhysRevLett.79.3306}.

\bibitem{Lidar}
Lidar DA.
\newblock Review of Decoherence Free Subspaces, Noiseless Subsystems, and
  Dynamical Decoupling.
\newblock http://arxivorg/abs/12085791. 2013;.

\bibitem{DiFrancesco}
Di~Francesco P, Mathieu P, S{\'e}n{\'e}chal D.
\newblock Conformal Field Theory.
\newblock Springer; 1997.

\bibitem{KZ}
Knizhnik VG, Zamolodchikov AB.
\newblock Current algebra and Wess-Zumino model in two dimensions.
\newblock Nuclear Physics B. 1984;247(1):83 -- 103.

\bibitem{Babujian}
Babujian HM.
\newblock Off-shell Bethe ansatz equations and N-point correlators in the SU(2)
  WZNW theory.
\newblock Journal of Physics A: Mathematical and General. 1993;26(23):6981.

\bibitem{Schechtman}
Schechtman VV, Varchenko AN.
\newblock Hypergeometric solutions of Knizhnik-Zamolodchikov equations.
\newblock Letters in Mathematical Physics. 1990;20(4):279--283.

\bibitem{Nielsen1}
Nielsen AEB, Cirac JI, Sierra G.
\newblock Quantum spin Hamiltonians for the SU (2) k WZW model.
\newblock Journal of Statistical Mechanics: Theory and Experiment.
  2011;2011(11):P11014.

\bibitem{Nielsen2}
Nielsen AEB, Cirac JI, Sierra G.
\newblock Laughlin Spin-Liquid States on Lattices Obtained from Conformal Field
  Theory.
\newblock Physical Review Letters. 2012;108(25).

\bibitem{Whittaker}
Whittaker ET, Watson GN.
\newblock A Course of Modern Analysis.
\newblock 4th ed. Cambridge University Press; 1996.

\bibitem{Lima-Santos}
Lima-Santos A, Utiel W.
\newblock Gaudin magnet with impurity and its generalized
  Knizhnik-Zamolodchikov equation.
\newblock International Journal of Modern Physics B. 2006;20(15):2175--2187.

\bibitem{Aomoto}
Aomoto K, Kita M.
\newblock Theory of Hypergeometric Functions.
\newblock Springer; 2011.

\bibitem{love}
Yoshida M.
\newblock Hyper Geometric Functions, My Love: Modular Interpretations of
  Configuration Spaces.
\newblock Friedrick Vieweg \& Son; 1997.

\bibitem{Reshetikhin}
Reshetikhin N, Varchenko A.
\newblock Quasiclassical Asymptotics of Solutions to the KZ Equations.
\newblock Geometry, topology, \& physics for Raoul Bott. 1995;p. 293--322.

\end{thebibliography}

\end{document}